\documentclass[twocolumn,showpacs,preprintnumbers,amsmath,aps,amssymb]{revtex4}
\usepackage{graphicx}
\usepackage{dcolumn}
\usepackage{bm}

\begin{document}

\title{Towards noncommutative supersymmetric quantum cosmology}
\author{W. Guzm\'an$^1$}
\email{wguzman@fisica.ugto.mx}
\author{M. Sabido$^1$}
\email{msabido@fisica.ugto.mx}
\author{J. Socorro$^{1,2}$}
\email{socorro@fisica.ugto.mx}
\affiliation{$^1$ Instituto de F\'{\i}sica de la Universidad de 
Guanajuato,\\
A.P. E-143, C.P. 37150, Le\'on, Guanajuato, M\'exico\\
$^2$ Facultad de Ciencias de la Universidad Aut\'onoma del Estado de 
M\'exico,\\
Instituto Literario  No. 100, Toluca, C.P. 50000, Edo de Mex, M\'exico}%

\begin{abstract}
Using the factorization approach of quantum mechanics, we obtain a family of
iso-spectral scalar potentials for noncommutative  quantum cosmology. The 
family we build is based
on a scattering Wheeler-DeWitt solution for the potential 
$V(\phi)=V_0e^{-\lambda\phi}$. We analyze the effects of noncommutativity on the iso-potentials and 
the possible relationship between noncommutativity and dark energy.
  \end{abstract}

\maketitle

{ The first venture into  quantum cosmology was engaged in the 1970's, through canonical 
quantization of minisuperspace models \cite{ryan} in which the gravitational and matter 
variables have been reduced to a finite number of degrees of freedom. 
The interest in the field was rekindled in the 80's by Hawking \cite{hawk}, emphasizing the path 
integral approach, reviving the interest in minisuperspace quantization. Recently, loop 
quantum cosmology has given new life to minisuperspace models, and as in the original proposal,
 the objective is to get some insight into the complete quantum theory of gravity by studying some simplified models.

Supersymmetry,  solves many  problems, such as the hierarchy problem,  or the dark matter conundrum. 
If one believes that supersymmetry is a fundamental property of nature, then supergravity should 
be the fundamental theory of gravity. This line of reasoning applied to cosmology gave birth to 
what is known as supersymmetric quantum cosmology (SUSY-QC)\cite{abue}. 
Several approaches to SUSY-QC were developed during the 90's \cite{90s}, all of them with their respective advantages 
and problems, yet one of the simplest approaches is the use of susy quantum mechanics to QC.

Since the beginning of this century, there has been a lot of interest in the old idea of noncommutative 
space-time \cite{snyder}. This renewed interest is a consequence
of the developments in string theory, where a noncommutative  gauge theory  action \cite{ref3,Douglas:2001ba} appears in a natural 
way.  The current formulations of gravity in noncommutative space-time 
\cite{ncg,wess} are highly nonlinear. Directly solving the cosmological equations from noncommutative gravity 
is a daunting task. In the last few years there have been several attempts to study the possible 
effects of noncommutativity in the cosmological
scenario \cite{romero, brand}. In particular, in \cite{ncqc}, the authors, analyzed the effects 
of noncommutativity in quantum cosmology by deforming the minisuperspace in a similar way as in 
noncommutative quantum mechanics \cite{gamboa1,jabbari}. This is achieved by introducing the Moyal 
product of functions in the Wheeler-DeWitt (WDW) equation, cunningly avoiding the difficulties of analyzing noncommutative cosmology.

The goal of this short paper is to use the methods of SUSY-QM, which can be considered an equivalent formulation of the Darboux
transformation method \cite{darboux}, that is  well-known in mathematics, and apply them to noncommutative quantum cosmology (NCQC). This is done
by introducing noncommutativity in the minisuperspace of the quantum 
cosmological model, and by using SUSY-QM
we get the iso-spectral wave functions as well as the iso-potentials for 
noncommutative quantum cosmology. This ideas have recently been applied in SUSY-QM \cite{gamboa}.}

With this in mind, we start with a flat,  homogeneous and isotropic
universe, { the  Friedmann-Robertson-Walker (FRW) metric
\begin{equation}
ds^2= -N^2(t) dt^2 + e^{2\alpha(t)}\left[dr^2
+r^2 d\Omega^2 \right], \, 
\label{frw}
\end{equation}
where $ a(t)=e^{\alpha(t)}$ is the scale factor, $\rm N(t)$ is the 
lapse function. From the Einstein-Hilbert action, 
coupled to a scalar field $\rm \phi$  with  scalar potential
$\rm V(\phi) = V_0 e^{-\lambda\phi}$,
\begin{equation}
 S =\int  dx^4 \sqrt{-g} \left[ R + 
 \frac{1}{2}g^{\mu \nu}\partial_\mu \phi \partial_\nu \phi+ 
 V(\phi)
  \right] \, ,
\label{accion}
\end{equation}
 we can write the  canonical Hamiltonian by means of the Legendre 
transformation 
and arrive to the FRW Hamiltonian
\begin{equation}
\rm {\cal H}=\frac{N}{12} e^{-3\alpha} \left[
\Pi_\alpha^2 - 6\Pi_\phi^2 - 12  e^{6 \alpha}  V(\phi)
\right], \label{uno}
\end{equation} 
where we have used the units $8\pi G=1$. In order to simplify the calculations from now on we will be working in 
the gauge $N=12e^{3\alpha}$, this will simplify the noncommutative 
formulation and because of the reparametrization  invariance of the theory,  
the physical implications are independent of the chosen gauge. Also in order to simplify the 
calculations we make the canonical transformation \cite{Guzman:2007zza}
\begin{eqnarray}
\rm x&=&\rm -6\alpha+\lambda\phi,\qquad~~ 
\rm \Pi_x= \frac{1}{\lambda^2-6} \left(\Pi_\alpha +\lambda\Pi_\phi\right), \label{cano}\\
y&=&-\sqrt{6}\lambda\alpha+\sqrt{6}\phi,
~~
\Pi_y=\frac{1}{\sqrt{6}(6-\lambda^2)} \left(\lambda\Pi_\alpha 
+6\Pi_\phi\right).\nonumber
\label{trans}
\end{eqnarray}
From this point forward our analysis will be done in this new set of minisuperspace variables, 
the advantages of this selection has been analyzed in 
\cite{Guzman:2007zza}.
The classical  Hamiltonian has the simple form
 ${\cal 
H}=-\beta \Pi^2_x+\beta \Pi^2_y-12V_0e^{-x} \approx 0$,
where $\beta$ is defined as $\beta\equiv6(\lambda^2-6)$. The usual canonical quantization is done by the usual 
identifications $\rm \Pi_{q^\mu}$=$\rm -i \partial_{q^\mu}$, and
we arrive to the WDW equation
 \begin{equation}
\rm  \frac{\partial^2 \Psi}{\partial x^2}
 - \frac{\partial^2 \Psi}{\partial y^2} + \frac{2 V_0}{6-\lambda^2 }
 e^{-x} \Psi=0,
\label{modified}
\end{equation}
in this formalism $\Psi$ is called the wave function of the universe. 

The proposal to introduce the noncommutative minisuperspace deformation  is achieved by introducing the following commutation 
relation between the minisuperspace variables
\begin{equation}
[\rm x,\rm y]=i\theta,
\label{ncms}
\end{equation}
this can be seen as an effective noncommutativity that could arise from a fundamental 
noncommutative theory of gravity. For example, if we start with the Lagrangian derived in \cite{ncg}, 
the noncommutative fields are a consequence of noncommutativity among the coordinates and 
then the minisuperspace variables would inherit  some effective noncommutativity. 
This we assume to be encoded in (\ref{ncms}), otherwise we would have a very complicated Hamiltonian for the higher order Lagrangian.

This effective noncommutativity can be formulated in terms of product of functions of the mini-superspace variables, 
with the Moyal star product of functions. The noncommutative WDW (NCWDW) equation is 
obtained by replacing  the products of functions by  star products.
We can show \cite{jabbari} that the effects of the Moyal star product are 
reflected only in a shift in the potential 
$V(\rm x,\rm y)\star\Psi(\rm x,\rm y)=V(\rm x+\frac{\theta}{2}\Pi_{\rm }y,\rm y-\frac{\theta}{2}\Pi_{\rm x})\Psi(x,y)$. 
Taking this in to account,  we arrive to
\begin{equation}
\rm  \left(\frac{\partial^2 }{\partial x^2}
- \frac{\partial^2}{\partial y^2} +\gamma
e^{-\left (x-\frac{i}{2}\theta\frac{\partial}{\partial \overline{y}}\right 
)}\right) \Psi=0,
\label{ncwdw}
\end{equation}
with $\gamma\equiv \frac{2V_0}{6-\lambda^2}$. 

Using the the ansatz $\Psi(x,y)=e^{\pm i\eta y} \Psi_x(x)$ and the 
property $\rm e^{i\theta \frac{\partial}{\partial y}} e^{\eta x}=e^{i\eta \theta}e^{\eta x}$
we arrive to
\begin{equation}
\frac{d^2 \Psi_x}{d^2 x^2} +\left( \gamma e^{-(x+\pm \frac{\theta}{2}\eta)} + \eta^2
\right) \Psi_x=0, \label{wdw3}
\end{equation}
 equation over we employ the susy  isospectral method,
who solution for $\lambda^2<6$ become
 \begin{equation}
\begin{split}
\Psi(x,y)=e^{\pm i\eta y}&\left[J_{2 \eta}\left(2 
\sqrt{\gamma} e^{ -(\frac{x}{2} \pm \frac{\theta}{4}\eta)}\right)
\right. \\
&+\left .J_{-2\eta}\left(2 
\sqrt{\gamma}
e^{-(\frac{x}{2} \pm \frac{\theta}{4}\eta)}\right)\right],
\end{split}
\end{equation}
this of course in the minisuperspace variables $x$ and $y$, which reduces to 
the corresponding commutative solution for $\theta=0$. When writing down the noncommutative WDW equation, 
one expects  additional ordering ambiguities to arise, for the example we are working, 
this can be ignored due to simple form of the potential 
{ and the classical canonical transformation we are using}.

As already mentioned, the goal of this letter is to apply the factorization 
approach of supersymmetric quantum mechanics  \cite{darboux},  to
  noncommutative quantum cosmology (for a full review of these techniques see \cite{cooper}). \\
For this we start with the WDW equation (or the Schroedinger equation if we are working in quantum mechanics),
 and proceed to find first  order  differential operators
that factorize the hamiltonian. The factorization technique is based on the fact that once we have a solution to the 
original WDW equation a super-potential function can be constructed, and a whole family of potentials can be found, 
with the particularity that all have the same energy spectrum \cite{ro-so}. These potentials are known as the iso-potentials. 
The family of iso-potentials can be parametrized and in the limit when the parameter goes to infinity we return to the original problem.
Due to simple modification on the noncommutative WDW equation with respect to its commutative counterpart 
factorizing the noncommutative WDW equation Eq.(\ref{wdw3})
immediately yields the iso-potentials and the iso-wavefunctions.

The total WDW isospectral wave function has the {generic}  form \cite{ro-so}
\begin{equation}
\rm \Psi_{iso}(x, y;\tau)=  \frac{ g(\tau)\,\Psi(x,y)}{{\cal I} + \tau},
\end{equation}
where $\rm g(\tau)=\sqrt{\tau(\tau+1)}$, $\tau$ is a continuous free parameter time-like and the function $\rm {\cal I}$ become 
\begin{equation}
\rm {\cal I}(x)=\int_0^x [\Psi_x(t)]^2 dt,
\end{equation}
where  the effects of the mini-superspace noncommutativity are incorporated on the 
Bessel functions. { The iso-potentials
are constructed from the noncommutative wave functions, so they will also be influenced by the 
noncommutative parameter $\theta$, and although  a closed expression for the noncommutative iso-potentials is difficult to construct, 
some general features can be seen 
 in  figures \ref{fig:esc1} and  \ref{fig:esc3}.}
\begin{figure}
\includegraphics[width=8cm]{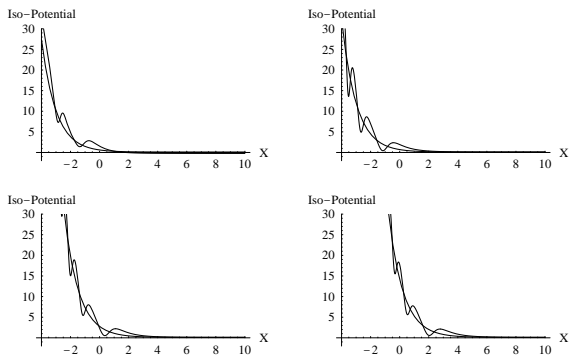}
\caption{\label{fig:esc1} Plot of the iso-potentials. These 
 plots correspond to values $\eta$= $\frac{1}{3}$, $\lambda=\sqrt{2}$, $\tau=.01$, and increasing
values of the noncommutative parameter $\theta=0,1,10,20$.  }
\end{figure}

We can see from (Fig.\ref{fig:esc1}) the effects of noncommutativity on the iso-potentials. As the value of $\theta$ is 
increased the  number of local minima and maxima  on the potential also increases. This is reminiscent to the effects of 
the noncommutative parameter on the probability density  in NC-QC, where new maxima and minima appear as the value of $\theta$ is 
increased; this can be interpreted in the framework of QC as new less probable universes to which our current universe can tunnel \cite{ncqc}.
\begin{figure}
\includegraphics[width=8cm]{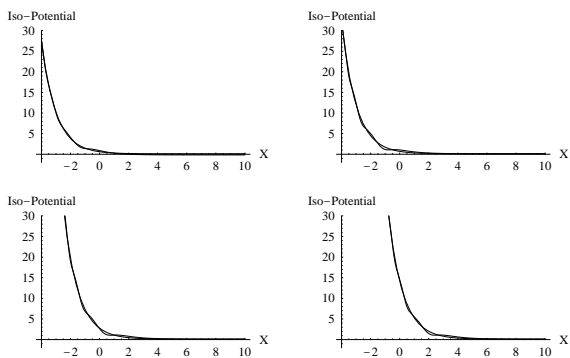}
\caption{\label{fig:esc3} 
Plot of the iso-potentials. These 
 plots correspond to values $\eta$= $\frac{1}{3}$, $\lambda=\sqrt{2}$, $\tau=.4$, and increasing
values of the noncommutative parameter $\theta=0,1,10,20$. }
\end{figure}
Another interesting effect of noncommutativity seems to be related to the value of the cosmological constant. 
The value of the minimum of the potential,  is related to the value of $\Lambda$, from the plot we can see that the value of the potential 
at its local minima can be adjusted by changing the value of $\theta$, from which one can expect a relationship 
between noncommutativity and the value of $\Lambda$.  
This of course is speculative, and further 
research with more realistic models is needed. But being $\Lambda$ related to the vacuum energy, 
one may be tempted to think that is value could be altered if we introduce a 
noncommutative deformation of space time (or in our case, on the minisuperspace). This kind of reasoning has already 
been proposed in \cite{lambda}, in that paper the author find a relationship between the noncommutative parameter and the 
cosmological constant and alleviate the discrepancy between the observed and calculated energy density.
With respect to (Fig.\ref{fig:esc3}), we see that as we increased the value of $\tau$ the iso-potentials goes to the original one, 
this is a general feature of the family of iso-potentials \cite{smg}.

We may now search for some cosmological effects, as already mentioned, the noncommutative parameter may be related 
to $\Lambda$ and so fort to dark energy and the current acceleration of the universe. In order to study the acceleration of the 
scale factor of the universe in scalar field cosmology it is convenient to work in the slow roll approximation. 
The slow roll condition is encoded in the parameter  
$\epsilon=\frac{1}{2}(\frac{V'(\phi)}{V(\phi)})^2<<1$ (in units of $M_p=1$), when $\epsilon\ge1$ the slow roll condition 
is broken and the universe no longer accelerates. In particular for the exponential potential $V(\phi)=e^{-\lambda\phi}$ 
the slow roll parameter is $\epsilon=\frac{1}{2}\lambda^2$, this means that once we fix $\lambda$ to have a non accelerating 
universe there is no mechanism to start an inflationary epoch.
 
In order to simplify the analysis we write the slow roll parameter in terms of the potential as a function of $x$ instead of the scalar 
field $\phi$;  taking the form  
$\epsilon=\frac{1}{2}\lambda^2({V'(x)}/{V(x)}+b)^2$, where $b={6}({\lambda^2-6})^{-1}$, 
for the original exponential potential $\frac{V'(x)}{V(x)}=-\lambda^2 b$, so again, once we fix the value of $\lambda$ to have inflation 
there is no way to end it. 
 For the iso-spectral potentials the approximate relation holds $\frac{V_{iso}(\phi)}{V(\phi)}{\approx}\frac{V_{iso}(x)}{V(x)}$ 
 where $V_{iso}(x)$ is the iso-spectral potential, and again we may write the slow roll parameter in a similar manner, now in 
 contrast with the original potential, once we fix the value of $\lambda=\sqrt{2}$ for a non  inflating epoch, as we increase 
 the value of $x$  the ratio ${V'_{iso}(x)}/{V_{iso}(x)}$ can be as small as we want,  this can be seen  
 from the plots of the iso-potentials (Fig.\ref{fig:esc1}).  In particular near the critical points of $V(x)$, 
 $\frac{V_{iso}(x)}{V(x)}$ can be small enough, in order to satisfy the slow roll condition, and give an accelerating epoch for the universe.

In this letter we have applied the factorization technique in order to find noncommutative iso-potentials to 
the noncommutative WDW equation, the scenario we have used corresponds to  FRW cosmological 
model coupled to a scalar field. We speculate on the possible relationship between dark energy and 
the noncommutative parameter, of course more realistic models need to be constructed, but the possible relationship 
between noncommutativity and dark energy is very attractive. Another possible area to which these ideas can be applied is in the very early 
universe, particularly in connection to inflation where noncommutativity might play a relevant role \cite{Guzman:2007zza}, 
the reason being that the iso-spectral method gives a complete family of iso-potentials that might give better agreement to 
the observational data, research in this line of reasoning in being done and will be reported elsewhere.

Finally we believe that this technique can be applied to
other areas of physics where the factorization approach of QM is used, 
the procedure gives a new parameter that can be introduced in a straight forward way and that it might be used  to have 
a better phenomenological agreement.

\acknowledgments{ \noindent This work was partially supported by CONACYT 
grants 47641and  62253. DINPO 38/07 and PROMEP grants UGTO-CA-3, UGTO-PTC-085 and CONCYTEG grant 07-16-K662-062 A01.}


\begin{thebibliography}{35}
\bibitem{ryan} For reviews, see M. P. Ryan, {\it Hamiltonian Cosmology} (Speinger, Berlin, 1972); 
M. MacCallum, in {\it General Relativity: An Einstein Centenary Survey}, edited by S. Hawking and W. Israel 
(Cambridge University Press, Cambridge, England, 1979).
\bibitem{hawk} S. Hawking, Pontif. Acad. Sci. Varia {\bf 48}, 563 (1982); J. B. Hartle and S. W. Hawking, Phys. Rev. D {\bf 28}, 2960 (1983).
\bibitem{abue} A. Mac\'ias, O. Obreg\'on and M. Ryan, Class. Quantum Grav. {\bf4}, 1477 (1987).
\bibitem{90s} J. Bene and R. Graham, Phys. Rev. D {\bf 49}, 799 (1994); O. Obreg\'on, J. Socorro and J. Ben\'{\i}tez,
Phys. Rev. D {\bf 47}, 4471 (1993); V.I. Tkach, J.J. Rosales and O. Obreg\'on, Class. Quantum Grav.
{\bf 13}, 2349 (1996); P.D. D'Eath, {\it Supersymmetry quantum cosmology} (Cambridge University Press,
Cambridge, England, 1996); J. Socorro and E.R. Medina, Phys. Rev. D {\bf 61}, 087702 (2000).
\bibitem{snyder} H. Snyder, Phys. Rev. {\bf 71}, 38 (1947).
\bibitem{ref3}  N. Seiberg and E. Witten, {JHEP} {\bf 9909:032} (1999).
\bibitem{Douglas:2001ba}
  M.~R.~Douglas and N.~A.~Nekrasov,
  Rev.\ Mod.\ Phys.\  {\bf 73}, 977 (2001)
  [arXiv:hep-th/0106048].
   \bibitem{wess}P.~Aschieri, M.~Dimitrijevic, F.~Meyer and J.~Wess,
  Class.\ Quant.\ Grav.\  {\bf 23} (2006) 1883.

\bibitem{ncg} H.~Garcia-Compean, O.~Obregon, C.~Ramirez and 
 M.~Sabido,
  Phys.\ Rev.\ D {\bf 68}, 044015 (2003); H.~Garcia-Compean, O.~Obregon, 
 C.~Ramirez and M.~Sabido,
  Phys.\ Rev.\ D {\bf 68}, 045010 (2003); A.H. Chamseddine, J.\ Math.\ Phys.\  {\bf 44}, 2534 (2003); 
  J.W.      Moffat, {Phys. Lett.} B {\bf 491}, 345 (2000);  
  P.~Aschieri, C.~Blohmann, M.~Dimitrijevic, F.~Meyer, P.~Schupp and J.~Wess,
  Class.\ Quant.\ Grav.\  {\bf 22}, 3511 (2005); L.~Alvarez-Gaume, F.~Meyer and M.~A.~Vazquez-Mozo,
  Nucl.\ Phys.\  B {\bf 753}, 92 (2006).

\bibitem{romero}
J.~M.~Romero and J.~A.~Santiago,
  Mod.\ Phys.\ Lett.\ A {\bf 20} (2005) 781
  [arXiv:hep-th/0310266].
\bibitem{brand}
   R.~Brandenberger and P.~M.~Ho,
  Phys.\ Rev.\ D {\bf 66} (2002) 023517; Q.~G.~Huang and M.~Li,
  Nucl.\ Phys.\ B {\bf 713} (2005) 219; Q.~G.~Huang and M.~Li,
  JHEP {\bf 0306} (2003) 014;H.~Kim, G.~S.~Lee, H.~W.~Lee and 
Y.~S.~Myung,
  Phys.\ Rev.\ D {\bf 70} (2004) 043521;H.~Kim, G.~S.~Lee and 
Y.~S.~Myung,
  Mod.\ Phys.\ Lett.\ A {\bf 20} (2005) 271;D.~J.~Liu and X.~Z.~Li,
  Phys.\ Rev.\ D {\bf 70} (2004) 123504.
\bibitem{ncqc}
H.~Garcia-Compean, O.~Obregon and C.~Ramirez,
Phys.\ Rev.\ Lett.\  {\bf 88}, 161301 (2002).
   \bibitem{gamboa1}
  J.~Gamboa, M. Loewe and J. C. Rojas, Phys.\ Rev.\ D {\bf 64},067901.

  \bibitem{jabbari}
  M. Chaichian, M. M. Sheikh-Jabbari, and  A. Tureanu, Phys. \ Rev. Lett. 
{\bf 86}, 2716.
\bibitem{darboux}G. Darboux, C.R. Acad. Sci. (Paris) {\bf 94}, 1456 (1882).
\bibitem{gamboa} A.~Das, H.~Falomir, J.~Gamboa and F.~Mendez,
  arXiv:0809.1405 [hep-th].
\bibitem{Guzman:2007zza}
  W.~Guzman, M.~Sabido and J.~Socorro,
  Phys.\ Rev.\  D {\bf 76} (2007) 087302.
\bibitem{cooper}F. Cooper, A. Khare and U. Sukhatme, Phys. Rep. {\bf 251},
	267 (1995).
	\bibitem{ro-so}  H. Rosu and J. Socorro, Il Nuovo Cimento B {\bf 113}, 
	683-689 (1998), [gr-qc/9606030].

\bibitem{lambda} E.~Mena, O.~Obregon and M.~Sabido,
  arXiv:0802.3393 [hep-th].
  \bibitem{smg} J. Socorro, M.A. Reyes and F.A. Gelbert, Physics Lett. A {\bf 313}, 338 (2003)

	

\end{thebibliography}
\end{document}